\documentstyle[11pt,fullpage]{article}
\begin{document}

\begin{flushright}
UFIFT-HEP-97-20\\
  hep-th/9707048\\
\end{flushright}
\vskip 3.0cm

\renewcommand{\thefootnote}{\fnsymbol{footnote}}
\def\footnoterule{\kern-3pt \hrule width \hsize \kern2.5pt}
\def\ie{{\it i.e.\ }}
\def\eg{{\it e.g.\ }}
\def\beq{\begin{equation}}
\def\eeq{\end{equation}}
\def\beqn{\begin{eqnarray}}
\def\eeqn{\end{eqnarray}}
\pagestyle{empty}

\begin{center}
{\Large\bf Supersymmetric Quantum Mechanics for String-Bits\footnote
{Supported in part by
the Department of Energy under grant DE-FG02-97ER-41029}
}

\vskip 2.5cm
\large{Charles B. Thorn}\footnote{E-mail  address: thorn@phys.ufl.edu}
\vskip 0.5cm
{\it Institute for Fundamental Theory\\
Department of Physics, University of Florida\\
Gainesville, FL 32611}
\end{center}
\vspace{1.2cm}
\begin{center}
{\bf ABSTRACT}
\end{center}
\noindent We develop possible versions of supersymmetric single
particle quantum mechanics, with application to
superstring-bit models in view. We focus principally on
space dimensions $d=1,2,4,8$, the transverse dimensionalities
of superstring in $3,4,6,10$ space-time dimensions. These are
the cases for which ``classical'' superstring makes sense, and also
the values of $d$ for which Hooke's force law is compatible with
the simplest superparticle dynamics. The basic question we address
is: When is it possible to replace such harmonic force laws with
more general ones, including forces which vanish at large distances?
This is an important question because forces between string-bits that
do not fall off with distance will almost certainly destroy
cluster decomposition. We show that the answer is affirmative for
$d=1,2$, negative for $d=8$, and so far inconclusive for $d=4$.
\noindent
\vfill
\newpage

\def\balpha{\mbox{\boldmath$\alpha$}}
\def\bgamma{\mbox{\boldmath$\gamma$}}
\def\bsigma{\mbox{\boldmath$\sigma$}}
\def\bepsilon{\hbox{\twelvembf\char\number 15}}
\def\Nlarge{N_c\rightarrow\infty}
\def\Tr{{\rm Tr}}
\newcommand{\ket}[1]{|#1\rangle}
\newcommand{\bra}[1]{\langle#1|}
\newcommand{\firstket}[1]{|#1)}
\newcommand{\firstbra}[1]{(#1|}
\setcounter{footnote}{0}
\pagestyle{plain}
\pagenumbering{arabic}
\renewcommand{\theequation}{\thesection.\arabic{equation}}

\section{Introduction}
\noindent In string-bit models\cite{thornmosc,thornrpa},
string is viewed as a polymer molecule, a bound system
of point-like constituents which enjoy a Galilei invariant
dynamics. This can be consistent with Poincar\'e invariant string, because
the Galilei invariance of string-bit dynamics is precisely that
of the transverse space of light-cone quantization. If the string-bit
description of string is correct, ordinary nonrelativistic many-body
quantum mechanics is the appropriate framework for string dynamics.
Of course, for superstring-bits, this quantum mechanics must be made
supersymmetric\cite{bergmantbits,bergmantoned}.

One virtue of the string-bit idea is that the forces which bind
the bits into string are also responsible for the interaction
between different parts of string. However, this double duty poses a
difficult challenge. It is no longer sufficient to posit
non-interacting ``free'' string as the fundamental degree
of freedom upon which interactions are imposed, for example,
by adding non quadratic terms to a ``string-field'' Lagrangian.
In order to be able to define a scattering matrix, the
composite polymers must obey cluster decomposition: the
interaction between widely separated polymers must vanish
sufficiently rapidly. Thus the simple expedient of modelling
the bond between nearest neighbors on a polymer by a harmonic
``spring'' force (Hooke's Law) will not do. Such a long-range
force would be indirectly felt between constituents in distant
polymers, violating clustering. The only forces in the fundamental
string-bit Hamiltonian should be of short range.

Supersymmetric quantum mechanics in one spatial dimension
is well known to allow an essentially arbitrary potential of
the form\cite{wittensusyqm}
\begin{equation}
V(x)=W^2(x)\pm W^\prime(x),
\end{equation}
where $W$ is the so-called superpotential.
The extension of supersymmetric quantum mechanics to higher dimensions
and to situations involving more than the minimal number of supercharges
has been explored by various authors\cite{crittenberg,flume,cooperk}.
These works address the problem of constructing superalgebras
with varying numbers of supercharges, and they conclude
that the possibilities are strikingly limited. For example,
within their working assumptions the only non-gauge
models with more than four supercharges require harmonic potentials.
An exception occurs when there is a gauge invariance
(that is when there are first class
constraints)
and the superalgebra is allowed
to close {\it modulo} a term which vanishes on gauge invariant states.
In the latter case the claim is that one is limited to those
quantum mechanical models that  result from dimensionally reducing
a supersymmetric Yang-Mills field theory\cite{crittenberg,flume}.
Our work does not conflict with any of these conclusions, but we
relax the assumption that the supercharges be linear
in the fermionic dynamical variables. At the same time we shall narrow
the field of possibilities by requiring
rotational invariance, which is not demanded in the
articles cited above.

The reason we want rotational invariance is that we regard
the one particle quantum mechanics we explore
in this paper as a sub-sector of a many body Galilei
invariant quantum
mechanics, {\it e.g.} the center of mass dynamics of a
two particle system. For the same reason, we retain another
restrictive assumption of \cite{crittenberg,flume,cooperk}, that
{\it the supercharges be linear in the particle momentum}. This will
then naturally implement the requirement from Galilei invariance
that each particle momentum enter the hamiltonian {\it quadratically}.
In this context it is worthwhile pointing out that without
some such restriction, the problem of exhibiting a superalgebra
with a highly nonharmonic hamiltonian is trivially solved.
Indeed, pick any hermitian operator $\Omega$ acting on the state space
of a (non-supersymmetric) quantum particle. Adjoin to the system
of quantum operators, spinor valued elements $S^A$ of a Clifford
algebra which commute with all other operators. Then $S^A\Omega$
satisfies the superalgebra with $H=\Omega^2/2$. Clearly there
is very little interesting about this supersymmetric
system: all we have is multiple copies of an original non-supersymmetric
system, with hamiltonian $\Omega^2/2$, whose only special feature is
a positive definite hamiltonian.

As mentioned above, the larger context of this work is to
find a satisfactory string-bit model for superstring in
which the forces between bits fall off with distance. In
earlier work\cite{bergmantgal} Bergman and I constructed
such a model with full Galilei supersymmetry, but it had
the fatal flaw that the potential
energy of the two bit system also vanished at large distance.
Since the model was supersymmetric this meant that any
bound state would have zero binding energy. In fact the
two-bit system did not possess even a zero energy bound state.
Thus we place particular emphasis on models in which
the forces vanish at large distance while the potential
energy approaches a non-vanishing (positive) constant. In such
models a supersymmetric ground state will automatically
have nonvanishing binding energy.

We shall only consider non-gauge models in this
article. Although we consider the problem of developing supersymmetric
particle quantum mechanics in arbitrary dimensionalities very
interesting, the fact that each dimensionality presents special features
leads us to narrow our attention to the dimensionalities of interest
for string theory: $d=1,2,4,8$, the transverse dimensionalities for
3, 4, 6, and 10 dimensional space-time. It is precisely for these
dimensionalities that it is possible to introduce a harmonic potential
into superparticle dynamics. Our main interest in this
article is to inquire whether short range forces are also possible for
these cases.
For $d=1$ Witten's analysis shows that they are.
We shall find that only harmonic forces are
possible in an 8-dimensional model with 8 supercharges but
short-range forces can be introduced in a
2-dimensional model with 2 supercharges. In the latter
case we shall exhibit supercharges with terms {\it cubic} in
the fermionic variables. This explicitly demonstrates that
the assumption of earlier authors that the supercharges
are linear in fermionic variables is overly restrictive.

The rest of the paper is organized as follows.
In Section 2 we discuss supersymmetry
in models with only harmonic forces. Although such models
have been studied before, we discuss them here
to set notation and also to display the supercharges
as {\it spinors} of the rotation group. In Section 3 we consider
generalizations to non-harmonic forces, and find the severe
restrictions mentioned above. The models we present
and study certainly
resemble some of those in the literature we have cited. There are,
however, new features that arise in some of our
models chiefly because,
motivated by superstring applications,  we
include more than the minimal number of fermionic
degrees of freedom. In Section 4 we analyze the
2-dimensional models in some detail, exhibiting cases
in which there is a supersymmetric ground state with
finite binding energy. Finally, we conclude
in Section 5 with a discussion of the application of our
results to superstring-bit models. We deal with the case $d=4$ in an
appendix.

\section{Harmonic Forces}
\label{sec2}
In the main text, we shall
limit our constructions of supersymmetry algebras
to those of hermitian supercharges $Q_A$ where $A$ is a
$2^{d/2}$ dimensional Majorana spinor index for $O(d)$. We also allow
a further restriction to a Majorana-Weyl
spinor when possible. Then $d$ is limited to those dimensions for which
the Dirac gamma matrices can be chosen real and symmetric:
\begin{equation}
\gamma^k_{AB}=\gamma^{k*}_{AB}=\gamma^k_{BA}
\qquad\{\gamma^k,\gamma^l\}=2\delta_{kl}
\label{symcliff}
\end{equation}
For application to superstring the interesting dimensions are
$d=2,8$\footnote{The other case interesting for superstring is
$d=4$ for which Majorana gamma matrices don't exist. Thus $Q_A$,
$S^A$, and ${\tilde S}^A$ are all non-hermitian.
This case is also consistent
with harmonic forces, provided a Weyl restriction is made. We
shall discuss this case in an appendix to keep the
line of argument uncluttered.}
(the transverse dimensionalities for string
in 4 and 10 dimensional space-time, respectively).
Similarly we choose Grassmann odd variables motivated by this
string application. We introduce a pair
of hermitian spinor valued Clifford
algebra elements $S^A,{\tilde S}^A$, satisfying
\begin{eqnarray}
\{S^A,S^B\}&=&2\delta_{AB}\qquad
\{{\tilde S}^A,{\tilde S}^B\}=2\delta_{AB}\nonumber\\
\{S^A,{\tilde S}^B\}&=&0.
\end{eqnarray}
Let $K$ be the dimensionality of the spinors. Then $K=2^{d/2}$ for
Dirac spinors and $K=2^{d/2-1}$ for Weyl spinors.

If the coordinates appear quadratically in the hamiltonian, they
should appear linearly in the supercharges. This motivates
the following {\it ansatz} for the supersymmetry generators
\begin{equation}
Q_A={\bf p}\cdot{\bgamma}_{AB}S^B+k{\bf x}\cdot{\bgamma}_{AB}{\tilde S}^B.
\label{harmq}
\end{equation}
By construction, the $Q_A$ transform as the components of a spinor under
rotations. A simple calculation yields
\begin{eqnarray}
\{Q_A,Q_B\}&=&2\delta_{AB}({\bf p}^2+k^2{\bf x}^2)\nonumber\\
& &{}+ik(\bgamma_{AC}\cdot
\bgamma_{BD}+\bgamma_{BC}\cdot\bgamma_{AD}){\tilde S}^CS^D
\end{eqnarray}
The supersymmetry algebra demands that this
anticommutator be $4\delta_{AB}H$.
If the r.h.s. happens to be proportional to the
Kronecker delta, the coefficient
will {\it define} the hamiltonian $H$.
To achieve this, we clearly need the
combination of gamma matrices to satisfy a special identity:
\begin{equation}
\bgamma_{AC}\cdot
\bgamma_{BD}+\bgamma_{BC}\cdot\bgamma_{AD}=2\delta_{AB}L_{CD}
\label{specialid}
\end{equation}
for some matrix $L$. In fact, simultaneous validity of
Eq.\ (\ref{symcliff}) and Eq.\ (\ref{specialid}) implies that
$L_{CD}=\delta_{CD}$ and $K=d$.  Clearly such
an identity will not hold generally,
but by using the Fierz identities one can test
whether it holds in various dimensions.
We shall examine it for $d=2,8$, the only cases for which $K=d$
and the Majorana representation is possible.

The Fierz identities use a complete set of spinor matrices to interchange
indices in the outer product of two matrices. The canonical
basis of spinor matrices is taken to be the anti-symmetrized products
of gamma matrices:
\begin{equation}
\gamma^{k_1k_2\cdots k_n}\equiv
\gamma^{[k_1}\gamma^{k_2}\cdots\gamma^{k_n]}
\end{equation}
where the square brackets denote the anti-symmetrized sum over all
permutations of the $n$ indices, normalized by dividing by $n!$. For
$d$ dimensions $n=0,1,2,\cdots d$, with the case $n=0$ understood as the
identity matrix. Then the
Fierz identity we need takes the form
\begin{equation}
\bgamma_{AC}\cdot\bgamma_{BD}=\sum_{n=0}^d C_n
\gamma^{k_1k_2\cdots k_n}_{AB}\gamma^{k_1k_2\cdots k_n}_{CD}
\end{equation}
Since the supersymmetry involves the l.h.s. symmetrized in $A,B$, the
desired result is obtained when the r.h.s. involves only
the identity and antisymmetric matrices. One easily finds
\begin{equation}
\gamma^{k_1k_2\cdots k_n T}=(-)^{n(n-1)/2}\gamma^{k_1k_2\cdots k_n}
\end{equation}
so the antisymmetric ones are $n=2,6,10,\ldots$ and $n=3,7,11,\ldots$.
Also $C_n$ is proportional to $d-2n$, so the term with $n=d/2$ is
never present.
Thus for $d=2$ only $n=0,2$ appear
in the Fierz identity and since the latter
is antisymmetric, the supersymmetry algebra closes by default. The
required identity Eq.\ (\ref{specialid}) assumes the form
\begin{equation}
d=2: \qquad \bgamma_{AC}\cdot
\bgamma_{BD}+\bgamma_{BC}\cdot\bgamma_{AD}=2\delta_{AB}\delta_{CD}
\label{specialid2}
\end{equation}

For $d=8$
the situation is a bit more complex:
all $n\neq4$ appear in the Fierz identity.
Of these $n=0,1,5,8$ are symmetric and contribute to the
anticommutator of supercharges. As is well known from the
Green-Schwarz formulation of the light-cone superstring, or from
the triality symmetry of $SO(8)$, the way to close the algebra, is
to make a Weyl restriction: with the index of $Q_A$ restricted to a
subset on which the chirality matrix
$\gamma_9\equiv\gamma^1\gamma^2\cdots\gamma^8$
is proportional to the identity. Then the $n=1,5$
terms won't enter the anticommutator (since they connect
indices with opposite values
of $\gamma_9$), and the $n=8$ term will simply
double the $n=0$ term. When the Weyl restriction is made, the
supercharges $Q$ will have chirality opposite to that of the spinors $S$.
The customary dotted index notation is then useful: the subset of
spinor indices $A$ with chirality $+1$ is denoted $a$ and the subset
with chirality $-1$ is denoted ${\dot a}$. For definiteness
we shall take the spinors $S$ to have undotted indices. Then the identity
Eq.\ (\ref{specialid}) takes the form
\begin{equation}
d=8: \qquad\quad\bgamma_{{\dot a}c}\cdot
\bgamma_{{\dot b}d}+\bgamma_{{\dot b}c}\cdot\bgamma_{{\dot a}d}=
2\delta_{{\dot a}{\dot b}}\delta_{cd}
\label{specialid8}
\end{equation}

In summary, we have reviewed the
extension of a quantum mechanical harmonic
oscillator to  a supersymmetric system for $d=2,8$. In doing so we
have limited our discussion to the Grassmann degrees of freedom suggested
by superstring and have assumed a minimal set of {\it hermitian}
spinor valued supercharges and the {\it ansatz} Eq.\ (\ref{harmq}). A similar
{\it ansatz} works for the case $d=4$ with non-hermitian supercharges.
In fact the way supersymmetry is
realized in this construction is essentially identical to
the way it is realized on a single mode of
superstring. The supersymmetry algebra does not close
for other dimensionalities if Eq.\ (\ref{harmq}) is assumed.
We do not
know to what extent constructions for other dimensionalities might
be made to work if, for example, the supercharges are
allowed to depend on the
spinor variables in a nonlinear way. In the following section we shall
find such nonlinearities are inevitable with non-harmonic forces, but
for harmonic forces, we have chosen not to investigate them.
%
\setcounter{equation}{0}
\section{Non-Harmonic Forces}
Although harmonic forces can provide useful models in
certain physical situations, their extreme long range character
threatens disastrous violations of cluster decomposition.
Thus the forces among string-bits should, at the very least,
vanish for large separations. In this section we examine the
possibilities for generalizing the supersymmetric quantum mechanics
models of the previous section to ones
with non-harmonic forces, including  those of short range.
Witten's one dimensional examples show that there is
no logical barrier to such a generalization. However, for
$d>1$ there are highly non-trivial constraints that
must be satisfied. Our efforts will only be fully successful for
$d=2$.

In light of the trivial realization of supersymmetry mentioned
in the introduction, to get dynamically interesting models
we must demand more than an algebraic realization of supersymmetry.
For example in supersymmetric field theory, locality provides
a powerful additional restriction on the dynamics. Locality is
not really applicable to one particle quantum mechanics. But an
analogously powerful restriction is provided by requiring the
particle momentum to enter the hamiltonian quadratically.
Thus it is reasonable to require that the
momentum dependence of the supercharges
in the presence of non-harmonic forces
be identical to that of the free or harmonic case:
\begin{equation}
Q_A={\bf p}\cdot{\bgamma}_{AB}S^B+{\hat Q}^A({\bf x}, S, {\tilde S}).
\label{qlinp}
\end{equation}
Requiring supersymmetry on this {\it ansatz} does indeed narrow the
possibilities drastically, as we shall see.

To begin write out the anticommutator,
\begin{eqnarray}
& &\{Q_A,Q_B\}=2{\bf p}^2\delta_{AB}+\{{\hat Q}_A,{\hat Q}_B\}\nonumber\\
& &\qquad\mbox{}+{1\over2}(\gamma^k_{AC}\{p^k,\{S^C,{\hat Q}_B\}\}
+\gamma^k_{BC}\{p^k,\{S^C,{\hat Q}_A\}\})\nonumber\\
& &\qquad\mbox{}+{i\over2}(\bgamma_{AC}\cdot\nabla[{\hat Q}_B,S^C]
+\bgamma_{BC}\cdot\nabla[{\hat Q}_A,S^C]),
\end{eqnarray}
and impose that each power of $p^k$ be separately proportional to a
$\delta_{AB}$. There are no constraints from the quadratic terms,
but the linear terms give
\begin{equation}
\gamma^k_{AC}\{S^C,{\hat Q}_B\}
+\gamma^k_{BC}\{S^C,{\hat Q}_A\}=2\Omega^k\delta_{AB},
\label{linconstr}
\end{equation}
while the momentum independent terms give
\begin{eqnarray}
\{{\hat Q}_A,{\hat Q}_B\}&+&
{i\over2}(\bgamma_{AC}\cdot\nabla[{\hat Q}_B,S^C]
+\bgamma_{BC}\cdot\nabla[{\hat Q}_A,S^C])\nonumber\\
&=&4V\delta_{AB}
\label{zeroconstr}
\end{eqnarray}
If these constraints can be satisfied the implied supersymmetric
hamiltonian would be
\begin{equation}
H={{\bf p}^2\over2}+{1\over4}(p^k\Omega^k+\Omega^kp^k)+V.
\end{equation}

We shall draw out the consequences of these constraints in stages.
First develop ${\hat Q}_A$ in an expansion in antisymmetrized monomials
of $S^A$:
\begin{equation}
{\hat Q}_A=\sum_{k=0}^{K}M^{B_1\cdots B_k}_A({\bf x},{\tilde S})
S^{[B_1}S^{B_2}\cdots S^{B_k]}.
\label{sexpand}
\end{equation}
Of course it is understood that $M$ is Grassmann odd (even) if $k$ is
even (odd). Without loss of generality we can take $M$ to be completely
antisymmetric in its upper indices.
The upper limit $K$ on the sum will be the spinor dimensionality,
$2^{d/2}$ for Majorana-Dirac spinors and $2^{(d-2)/2}$ for
Majorana-Weyl spinors. For the cases of particular interest to us
($d=2$ Dirac, $d=8$ Weyl) the upper limit is numerically equal to $d$.
Applying the constraints Eq.\ (\ref{linconstr}) leads to no restriction on
the the $k=0$ case and for $k>0$ amounts to:
\begin{equation}
\gamma^i_{AC}M_B^{CB_2\cdots B_k}+\gamma^i_{BC}M_A^{CB_2\cdots B_k}
=2\delta_{AB}{A}^i_{B_2\cdots B_k},
\label{lindev}
\end{equation}
where we have explicitly used the antisymmetry of $M$ in all its
upper indices and ${A}$ is as yet undetermined.
When Eq.\ (\ref{specialid2}) or Eq.\ (\ref{specialid8}) hold  (for us this
requires
$d=2,8$),
it is easy to solve
Eq.\ (\ref{lindev}) using them and  the identities Eq.\ (\ref{symcliff}):
simply put $B=A$ to get
\begin{equation}
\gamma^i_{AC}M_A^{CB_2\cdots B_k}={A}^i_{B_2\cdots B_k}.
\end{equation}
Then Eq.\ (\ref{specialid2}) with $A=B$ reads $\bgamma_{AC}\cdot
\bgamma_{AD}=\delta_{CD}$, so that
\begin{equation}
M_A^{CB_2\cdots B_k}=\gamma^i_{AC}{A}^i_{B_2\cdots B_k}.
\end{equation}
But for $k>1$ this factorized form is inconsistent with the
antisymmetry of $M$ in its upper indices: the r.h.s. must
vanish for $C=B_2$ which would imply ${A}^i_{B_2\cdots B_k}=0$.
This follows from the Clifford algebra Eq.\ (\ref{symcliff}).
Thus only the terms in Eq.\ (\ref{sexpand}) with $k=0,1$ contribute so
that the supercharges simplify to
\begin{equation}
{Q}_A=({\bf p}+{\bf A}({\bf x},{\tilde S}))\cdot\bgamma_{AB}S^{B}
+M_A({\bf x},{\tilde S}).
\label{simpleq1}
\end{equation}
Notice how ${\bf{A}}$ enters exactly as the vector potential of a
gauge field. The simplification so far is not so surprising: we
have begun with the restriction that $Q$ is linear in the components
of ${\bf p}$ and have found that the dependence on $S$, the spinor
naturally associated with the momentum, must also be linear. Although
the simple form Eq.\ (\ref{simpleq1}) has only been proved
to be forced when $d=2,8$, we
shall analyze it as an {\it ansatz} for generic $d$ in the following.

With the form of supercharges in Eq.\ (\ref{simpleq1}), the supersymmetry
algebra closes up to terms independent of momentum. Requiring
complete closure will put constraints on ${\bf A}$ and $M_A$. Again
it is efficient to organize the terms that arise from the
anticommutator according to powers of the spinor $S^B$:
\begin{eqnarray}
& &\{{Q}_A,{Q}_B\}=2({\bf p}+{\bf A})^2\delta_{AB}+\{M_A,M_B\}\nonumber\\
& &\qquad\mbox{}+([M_B,({\bf p}+{\bf A})\cdot\bgamma_{AC}]
               +[M_A,({\bf p}+{\bf A})\cdot\bgamma_{BC}])S^C\nonumber\\
& &\qquad\quad\mbox{} + [p^k+A^k,\ p^m+A^m]
                       \gamma^k_{AC}\gamma^m_{BD}S^{[C}S^{D]}.
\end{eqnarray}
Each power of $S$ must be separately proportional to $\delta_{AB}$.
Look first at the term quadratic in $S$. The Dirac matrices enter
in the combination
\begin{equation}
\gamma^k_{AC}\gamma^m_{BD}-\gamma^k_{AD}\gamma^m_{BC}
-\gamma^m_{AC}\gamma^k_{BD}+\gamma^m_{AD}\gamma^k_{BC}.
\label{gammacomb}
\end{equation}
In general dimensionality this combination will {\it not}
be proportional to $\delta_{AB}$, in which case closure of the
supersymmetry algebra will impose linear relations among the
components of $[p^k+A^k,\ p^m+A^m]\equiv iF^{mk}$. In sufficiently
high dimensionality, there will be so many independent
conditions to force the vanishing of all components:
\begin{eqnarray}
[p^k+A^k,\ p^m+A^m]&=&
i(\partial^mA^k-\partial^kA^m)
-[A^m,\ A^k]\nonumber\\
&\equiv& iF^{mk}
=0,\qquad{\rm generic~}d.
\label{fzero}
\end{eqnarray}
This constraint on $F^{mk}$ might be relaxed, partially or completely,
in specific dimensionalities. In view of rotational invariance a
partial relaxation is an option only in $d=4$, where self-duality
can provide a rotationally invariant linear relation.

The generic constraints Eq.\ (\ref{fzero}) are so powerful, that it is
worth pursuing their consequences once and for all.
Thinking of $F$ as a nonabelian field strength shows us immediately
that the solution of this constraint is that ${\bf A}$ is a
``pure gauge''
\begin{equation}
{\bf A}=\Omega^\dagger i\nabla\Omega,\quad\Omega^\dagger\Omega=I,
 \quad\qquad{\rm generic~}d.
\end{equation}
But with ${\bf A}$ of this form, the supercharges are unitary equivalents
of charges with ${\bf A}=0$. Thus without loss of generality we can
take
\begin{equation}
{Q}_A={\bf p}\cdot\bgamma_{AB}S^{B}+M_A({\bf x},{\tilde S}),
\quad\qquad{\rm generic~}d.
\label{simpleq2}
\end{equation}
With closure conditions
\begin{eqnarray}
\{M_A,M_B\}&=& 4V\delta_{AB}\label{genclosure0}\\
({\bf\nabla}M_B\cdot\bgamma_{AC}
               +{\bf \nabla}M_A\cdot\bgamma_{BC})&=&2\Psi_C\delta_{AB}.
\label{genclosure1}
\end{eqnarray}

The second closure condition Eq.\ (\ref{genclosure1}) can be inverted
by setting $B=A$, multiplying both sides by $\bgamma_{AC}$,
summing over $C$, and using the Clifford algebra:
\begin{equation}
\nabla M_A=\bgamma_{AC}\Psi_C.
\end{equation}
Note that this form satisfies Eq.\ (\ref{genclosure1}) only
if Eq.\ (\ref{specialid2}) or Eq.\ (\ref{specialid8}) hold, \ie only
if $d=2,8$.
The integrability condition for the last displayed equation is
\begin{equation}
(\nabla^i\gamma^j-\nabla^j\gamma^i)\Psi=0,
\end{equation}
where matrix multiplication is understood. For fixed {\it distinct}
$i,j$, multiply this equation on the left by the matrix
$\gamma^i\gamma^j=-\gamma^j\gamma^i$; this leads to
\begin{equation}
(\nabla^{i}\gamma^i+\nabla^{j}\gamma^j)
\Psi=0\qquad{\rm each~distinct~pair~} i,j.
\end{equation}
For $d>2$ this in turn implies that $\nabla_i\gamma_i\Psi_C=0$
for each $i$, and,
since $\gamma^i$ is an invertible matrix,
$\Psi_C$ is {\it independent} of the coordinates ${\bf x}$. So we
conclude that Eq.\ (\ref{genclosure1}) holds for $d>2$ if and only if
\begin{equation}
M_A=M^0_A({\tilde S})+{\bf x}\cdot\bgamma_{AC}\Psi_C({\tilde S}).
\label{mlinx}
\end{equation}
For $d=2$ the integrability condition is less stringent, implying
only that $\Psi\equiv\Psi_1-i\Psi_2$ is a holomorphic function of
$z=x_1+ix_2$. But rotational invariance (see the next section),
{\it i.e.} that $\Psi_C$
transform as a spinor with spin $\pm1/2$, is enough
to force the linear dependence on coordinates Eq.\ (\ref{mlinx}) for
$d=2$ as well.
So we are nearly back to the harmonic force case of the previous
section. It only remains to impose the other closure condition
Eq.\ (\ref{genclosure0}). Look first at the term in the anticommutator
quadratic in ${\bf x}$, which yields
\begin{equation}
(\gamma^m_{AC}\gamma^n_{BD}+\gamma^n_{AC}\gamma^m_{BD})\{\Psi_C,\Psi_D\}
=4\delta_{AB}V^{mn}.
\end{equation}
by setting $B=A$, summing over $A$, and using the Clifford algebra, we
easily see that $V^{mn}\propto\delta_{mn}$
so put  $V^{mn}=V_2\delta_{mn}$. Then setting
$B=A$ (but not summing) and inverting in the now familiar way, we find
\begin{equation}
\{\Psi_C,\Psi_D\}
=2V_2\delta_{mn}\gamma^m_{AC}\gamma^n_{AD}
=2V_2\delta_{CD}.
\end{equation}
It follows that $V_2$ commutes with $\Psi_C$.
Thus $\Psi_C/\sqrt{V_2}$
which is a function only of ${\tilde S}$
is a spinor whose components satisfy a Clifford algebra isomorphic
to that ${\tilde S}_C$. Hence it is unitarily equivalent to the latter
and $V_2\equiv k^2$ is a positive $c$-number. Thus
we can identify $\Psi_C$ with $k{\tilde S}$ and write
\begin{equation}
M_A=M^0_A({\tilde S})+k{\bf x}\cdot\bgamma_{AC}{\tilde S}_C.
\end{equation}
But now an identical argument to that which led from Eq.\ (\ref{qlinp}) to
Eq.\ (\ref{simpleq1}) shows that
$M^0_A\propto{\bf v}\cdot\bgamma_{AC}{\tilde S}^C$,
where ${\bf v}$ is a $c$-number vector. Since we are assuming rotational
invariance this vector must vanish. We then conclude that for generic
dimensionality, the {\it ansatz} of Eq.\ (\ref{simpleq1}), which is a
logical consequece of Eq.\ (\ref{qlinp}) for $d=2,8$, leads inevitably
to the supersymmetric  harmonic oscillator discussed
in the previous section.

There remains the loophole that in certain specific dimensionalities the
combination of Dirac gamma matrices Eq.\ (\ref{gammacomb}) might fortuitously
be proportional to $\delta_{AB}$, or it might not have enough independent
components to force {\it all} components of $F$ to vanish.
Here we only consider the cases $d=2,8$. The case of $d=4$ where
self-duality is a possibility is treated in the appendix.
First develop a Fierz-like expansion of Eq.\ (\ref{gammacomb}):
\begin{eqnarray}
& &\gamma^k_{AC}\gamma^m_{BD}-\gamma^k_{AD}\gamma^m_{BC}
-\gamma^m_{AC}\gamma^k_{BD}+\gamma^m_{AD}\gamma^k_{BC}\nonumber\\
& &\qquad\qquad\qquad\qquad=\sum_{n=0}^d C^{k_1k_2\cdots k_n}_{CDkm}
\gamma^{k_1k_2\cdots k_n}_{AB}.
\label{gammacombfierz}
\end{eqnarray}
Because the l.h.s. is symmetric in $AB$, only the terms with
$n=0,4,8,\dots$ and $n=1,5,\cdots$ will enter the sum on the
r.h.s.

For $d=8$ we have made the Weyl restriction and have
agreed to take $AB\rightarrow{\dot a}{\dot b}$ to be dotted and
$CD\rightarrow cd$ to be undotted. Then the terms with $n$
odd will not appear and those  with $n>4$
give nothing new. Thus we are limited to the terms with
$n=0,4$. Moreover, since the l.h.s. is antisymmetric
in $cd$ the coefficient $C^{k_1k_2\cdots k_n}_{cdkm}$ must
be a linear combination of the matrices $\gamma^{l_1l_2}_{cd}$.
By rotational invariance, the expansion simplifies to
\begin{eqnarray}
& &\gamma^k_{{\dot a}c}\gamma^m_{{\dot b}d}
-\gamma^k_{{\dot a}d}\gamma^m_{{\dot b}c}
-\gamma^m_{{\dot a}c}\gamma^k_{{\dot b}d}
+\gamma^m_{{\dot a}d}\gamma^k_{{\dot b}c}\nonumber\\
& &\qquad\quad=C_0\delta_{{\dot a}{\dot b}}\gamma^{km}_{cd}+C_4
\gamma^{kmk_1k_2}_{{\dot a}{\dot b}}\gamma^{k_1k_2}_{cd}
\qquad d=8.
\end{eqnarray}
By tracing the indices ${\dot a}{\dot b}$, one easily finds
that $C_0=1/2$. Then multiplying by $\gamma^{r_1r_2}_{dc}$
and summing over $c,d$ allows one to conclude that $C_4=1/4\neq0$.
There are 35 independent components of
$\gamma^{klmn}_{\dot a\dot b}$, which is
more than enough to force all
28 components of $F$ to vanish. Specifically, supersymmetry requires
\begin{equation}
F^{km}\gamma^{kmk_1k_2}_{{\dot a}{\dot b}}\gamma^{k_1k_2}_{cd}=0
\end{equation}
for all $c,d,\dot a, \dot b$. Multiplying the l.h.s. by
$\gamma^{l_1l_2l_3l_4}_{{\dot b}{\dot a}}\gamma^{m_1m_2}_{dc}$,
and summing over the repeated subscripts, reexpresses this condition as
\begin{equation}
F^{km}(\delta^k_{[l_1}\delta^m_{l_2}\delta^{k_1}_{l_3}\delta^{k_2}_{l_4]}
+\epsilon^{kmk_1k_2l_1l_2l_3l_4})=0.
\end{equation}
Choosing $k_1,k_2,l_1,l_2,l_3,l_4$ all different singles out a unique
component of $F$, so
for $d=8$ we conclude that,
within the {\it ansatz} Eq.\ (\ref{qlinp}), only harmonic forces are consistent
with the supersymmetry algebra.

For $d=2$, there is no Weyl restriction, and both  $n=0,1$ could
appear  in the expansion. However, the antisymmetry in $CD$ means
that the $CD$ dependence must be carried by $\gamma^{l_1l_2}_{CD}
=\epsilon_{l_1l_2}(\gamma^1\gamma^2)_{CD}$. Thus rotational
invariance excludes the presence of $n=1$. We find
\begin{eqnarray}
& &\gamma^k_{AC}\gamma^m_{BD}-\gamma^k_{AD}\gamma^m_{BC}
-\gamma^m_{AC}\gamma^k_{BD}+\gamma^m_{AD}\gamma^k_{BC}\nonumber\\
& &\qquad\qquad\qquad\quad=2\delta_{AB}\epsilon^{km}(\gamma^1\gamma^2)_{CD}
\qquad d=2.
\end{eqnarray}
Thus ${\bf A}$ is not constrained to be a pure gauge, and the
closure condition Eq.\ (\ref{genclosure1}) is relaxed to
\begin{eqnarray}
({\bf\nabla}M_B-i[M_B,{\bf A}])\cdot\bgamma_{AC}
               &+&({\bf \nabla}M_A-i[M_A,{\bf A}])\cdot\bgamma_{BC}
\nonumber\\
&=&2\Psi_C\delta_{AB}
\label{clos2}
\end{eqnarray}
In the next section we shall find that these relaxed constraints
leave room for an essentially arbitrary rotationally invariant
potential.
\setcounter{equation}{0}
\section{Supersymmetric Quantum Mechanics in 2 Dimensions.}
\label{sec4}
In two dimensions, the two component spinor
${\tilde S}$ enters $M$ at most linearly
and can enter ${\bf A}$ at most quadratically. This makes a
general analysis of this case mercifully tractable. Thus $M_A$
can be assumed to be, putting $\Gamma\equiv \gamma^1\gamma^2$, and
using rotational invariance,
\begin{eqnarray}
M_A&=&[({B}_1({\bf x}^2)+{B}_2({\bf x}^2)
\Gamma){\bf x}\cdot\bgamma]_{AC}{\tilde S}^C\nonumber\\
& &\qquad\qquad\mbox{}+({C}_1({\bf x}^2)+{C}_2({\bf x}^2)
\Gamma)_{AC}{\tilde S}^C
\end{eqnarray}
where the coefficients are all real. A short evaluation using
the closure condition Eq.\ (\ref{genclosure0}) then shows that either
$C_1=C_2=0$ or $B_1=B_2=0$. Furthermore a unitary
transformation of  the form
$\exp(\alpha({\bf x}^2)\Gamma_{CD}{\tilde S}^C{\tilde S}^D/4)$
can be used to rotate $B_2$ (or $C_2$) away, so we can assume
\begin{equation}
M_A=B({\bf x}^2){\bf x}\cdot\bgamma_{AC}{\tilde S}^C
\qquad{\rm or}\qquad M_A={C}({\bf x}^2){\tilde S}^A
\end{equation}
Similarly, rotational invariance restricts ${\bf A}$ to the form
\begin{equation}
{\bf A}={\bf A}_1({\bf x})+{1\over4}{\bf A}_2({\bf x})
i\Gamma_{CD}{\tilde S}^C{\tilde S}^D.
\end{equation}
It follows that
\begin{equation}
-i[{\tilde S}^A, {\bf A}]={\bf A}_2\Gamma_{AC}{\tilde S}^C,
\end{equation}
Then in the case that $C({\bf x}^2)$ is nonzero it is immediate
that Eq.\ (\ref{clos2}) cannot be satisfied unless $C$ is a
constant and ${\bf A}_2=0$. The other case with $B\neq0$
involves
\begin{eqnarray}
{\bf\nabla}M_B-i[M_B,{\bf A}]&=&(B({\bf x}^2)\bgamma_{BC}
+2{\bf x}B^\prime({\bf x}^2){\bf x}\cdot\bgamma_{BC}\nonumber\\
& &\mbox{}+B({\bf x}^2){\bf A}_2{\bf x}
\cdot(\bgamma\Gamma)_{BC}){\tilde S}^C.
\end{eqnarray}
This result allows a nontrivial solution
of Eq.\ (\ref{clos2}) because of the special properties of two
dimensions. The dual of a vector $v^i_D\equiv\epsilon^{ij}v^j$
is a vector. Furthermore $\bgamma\Gamma=\bgamma_D$ from which
follows ${\bf x}\cdot\bgamma\Gamma=-{\bf x}_D\cdot\bgamma$. Finally
we have the identity
\begin{equation}
v^i_Dv^j_D=\epsilon^{ik}v^k\epsilon^{jl}v^l=\delta_{ij}{\bf v}^2
-v^iv^j.
\end{equation}
Applying these special properties to the first term on the l.h.s
of Eq.\ (\ref{clos2}), leads to (with ${\bf A}_2=-2{\bf x}_DB^\prime/B$)
\begin{eqnarray}
&&({\bf\nabla}M_B-i[M_B,{\bf A}])\cdot\bgamma_{AC}=(B({\bf x}^2)
\bgamma_{BD}\cdot\bgamma_{AC}+\nonumber\\
&&~~2B^\prime({\bf x}^2){\bf x}\cdot\bgamma_{AC}{\bf x}\cdot\bgamma_{BD}-
B({\bf x}^2){\bf A}_2\cdot\bgamma_{AC}{\bf x}_D
\cdot\bgamma_{BD}){\tilde S}^D
\nonumber\\
&&\qquad\qquad= (B({\bf x}^2)+2{\bf x}^2B^\prime({\bf x}^2))
\bgamma_{BD}\cdot\bgamma_{AC}{\tilde S}^D,
\end{eqnarray}
Doing the same to the second term and making use of Eq.\ (\ref{specialid2})
shows that Eq.\ (\ref{clos2}) holds with
\begin{equation}
\Psi_C=(B({\bf x}^2)+2{\bf x}^2B^\prime({\bf x}^2)){\tilde S}^C.
\end{equation}
Putting all the pieces together, we conclude that for $d=2$ the
supersymmetry generators can be taken to be
\begin{eqnarray}
Q_A&=&\left({\bf p}+{\bf A}_1({\bf x})
-{\bf x}_D{B^\prime({\bf x}^2)\over2B({\bf x}^2)}
i\Gamma_{CD}{\tilde S}^C{\tilde S}^D\right)\cdot\bgamma_{AB}S^B
\nonumber\\
& &\qquad\qquad\qquad\qquad\mbox{}
+B({\bf x}^2){\bf x}\cdot\bgamma_{AC}{\tilde S}^C.
\label{2dq}
\end{eqnarray}
More general forms are unitarily equivalent to this. Incidentally, it
is apparent from Eq.\ (\ref{2dq}) that the case ${\bf A}=0$ does indeed
reduce to the supersymmetric oscillator, as we mentioned in the
previous section.

Before we analyze some of these two dimensional
models, it is worthwhile noting
a particularly simple way of understanding the  structure of
Eq.\ (\ref{2dq}).  Since $Q_A$ has two hermitian components, we
can combine them into a single non-hermitian supercharge,
with superalgebra:
\begin{eqnarray}
Q\equiv {1\over\sqrt2}(Q_1&+&iQ_2)\qquad Q^2=0\nonumber\\
\{Q,Q^\dagger\}&=&Q_1^2+Q_2^2=4H.
\end{eqnarray}
The last equation may be taken simply as the {\it definition} of
the hamiltonian $H$. Thus the only nontrivial content of
supersymmetry is the nilpotency of $Q$. If one nilpotent $Q_0$
can be found (for example, the supercharge for harmonic forces)
then $Q=YQ_0Y^{-1}$, where $Y$ is {\it any} invertible operator,
will also be nilpotent. Of course if $Y$ is unitary, the dynamics
is equivalent to that of the original system, and nothing new
is obtained. However, if $Y^\dagger\neq Y^{-1}$, one obtains
by this device a completely new system. It is easy to show that
the form Eq.\ (\ref{2dq}) is obtained in this way if $Y$ is restricted to
be a function of ${\bf x}$ and ${\tilde S}$.
That restriction is precisely what is
needed to implement our requirement that the momentum dependent
part of the supercharges be that of the harmonic or free system.

To facilitate the solution of these two dimensional models,
it is useful to introduce a representation for $S, {\tilde S}$
in terms of $4\times4$ matrices:
\begin{eqnarray}
S^1=\pmatrix{0&I\cr I&0\cr}
\qquad S^2=\pmatrix{0&i\sigma^3\cr -i\sigma^3&0}\nonumber\\
{\tilde S}^1=\pmatrix{0&i\sigma^1\cr -i\sigma^1&0}
\qquad {\tilde S}^2=\pmatrix{0&i\sigma^2\cr -i\sigma^2&0}
\label{4by4rep}
\end{eqnarray}
At the same time let's fix the representation of the $2\times2$
gamma matrices as
\begin{eqnarray}
\gamma^1=\pmatrix{0&1\cr1&0\cr}\qquad \gamma^2=\pmatrix{1&0\cr0&-1\cr}
\nonumber\\
\Gamma\equiv\gamma^1\gamma^2=\pmatrix{0&-1\cr1&0\cr}
\end{eqnarray}
It is also useful to identify the generator of rotations in the
plane in order to keep track of the quantum numbers of the energy
eigenstates. We find that the angular momentum is given by
\begin{equation}
J=x^1p^2-x^2p^1+{i\over4}(S^1S^2+{\tilde S^1}{\tilde S^2})
\end{equation}
In terms of the matrix representation Eq.\ (\ref{4by4rep}) the
angular momentum takes the forms
\begin{equation}
J=x^1p^2-x^2p^1-{1\over2}\pmatrix{0&0\cr0&\sigma^3},
\end{equation}
from which we see that the top two components of the four
component wave function have spin 0 while the bottom two
components have spin $\mp1/2$.

The matrix representations of the supercharges are
\begin{equation}
Q_A=\pmatrix{0&q_A^\dagger\cr q_A&0\cr}
\end{equation}
where
\begin{eqnarray}
q_1&=& p^2-ip^1\sigma^3
+{B^\prime\over B}(x^1\sigma^3+ix^2I)-iB(x^2\sigma^1
+x^1\sigma^2)\nonumber\\
q_2&=& p^1+ip^2\sigma^3
-{B^\prime\over B}(x^2\sigma^3-ix^1I)-iB(x^1\sigma^1
-x^2\sigma^2)\nonumber\\
 &=&i\sigma^3q_1
\end{eqnarray}
One fundamental question to ask of any supersymmetric system is
whether the ground state is supersymmetric, {\it i.e.} whether it
is annihilated by the $Q_A$. Any such state is automatically
a zero energy eigenstate of the Hamiltonian. For the case
of harmonic forces the ground state is supersymmetric
and has spin zero, so it is natural
to look for a supersymmetric state in the spin zero sector, so
we assume $\Psi=(\psi_1,\psi_2,0,0)$.
Denoting $\psi\equiv(\psi_1,\psi_2)$, we search for solutions of
\begin{eqnarray}
q_1\psi=[p^2-ip^1\sigma^3
&+&{B^\prime\over B}(x^1\sigma^3+ix^2I)\nonumber\\
&-&iB(x^2\sigma^1+x^1\sigma^2)]\psi=0.
\end{eqnarray}
Any such solution will automatically be annihilated by $q_2$. In terms of
components, this equation becomes the pair
\begin{eqnarray}
[p^2-ip^1+{B^\prime\over B}(x^1+ix^2)]
\psi_1-B[x^1+ix^2]\psi_2&=&0\nonumber\\{}
[p^2+ip^1-{B^\prime\over B}(x^1-ix^2)]\psi_2+B[x^1-ix^2]\psi_1&=&0.
\end{eqnarray}
These equations can be directly integrated in the case of zero
angular momentum, in which case we can assume that $\psi_{1,2}$ are
functions of ${\bf x}^2$, which implies $(p^2\mp ip^1)\psi_{1,2}
=\mp2(x^1\pm ix^2)\psi_{1,2}^\prime$, where the prime indicates
differentiation with respect to $u\equiv r^2$.
Then the equations reduce to
\begin{eqnarray}
-2\psi_1^\prime+{B^\prime\over B}\psi_1-B\psi_2&=&0\nonumber\\{}
2\psi_2^\prime-{B^\prime\over B}\psi_2+B\psi_1&=&0.
\end{eqnarray}
Taking sums and differences leads to two decoupled equations for
$\psi_\pm=\psi_1\pm\psi_2$, with solutions
\begin{equation}
\psi_\pm({\bf x}^2)=K_\pm\sqrt{B({\bf x}^2)\over B(0)}
\exp\left\{\mp\int_0^{{\bf x}^2}duB(u)\right\}.
\end{equation}
Clearly one or the other of these wave functions is normalizable
provided the integral
in the exponent diverges as $|{\bf x}|\to\infty$ sufficiently rapidly
and the sign $\mp$ is
chosen according to whether the integral blows up positively
or negatively. For a finite non-zero binding energy, the wave function
should fall off exponentially with distance, which would require
that $B(u)\sim 1/\sqrt{u}$ as $u\to+\infty$. The harmonic case
corresponds to $B(u)={\rm constant}$ and gaussian wave functions.

\setcounter{equation}{0}
\section{Applications to String-bit Models and Concluding Remarks}
We have managed to construct a supersymmetric one particle
quantum mechanics with a short-range potential ({\it i.e.} forces
vanishing at large distances) in 2 dimensions, but not in 8
dimensions. We hope this provides a useful step toward
a physically satisfactory string-bit model of superstring.
In an earlier work\cite{bergmantbits} Bergman and I constructed
a bit model of the free type IIB superstring based on a harmonic nearest
neighbor bond potential. The model possessed full Galilei
supersymmetry for
both $d=2$ and $d=8$. We can use the results of the preceding
sections to relax the restriction to harmonic forces in the
$d=2$ case.

In the ``bare polymer'' approximation ($N_c\to\infty$) the
string bit supercharges Ref.\cite{bergmantbits}, acting
on a single polymer state with $M$ bits, took the form:
\begin{equation}
Q^A=\sum_{k=1}^M[{\bf p}_k\cdot\bgamma_{AB}S_k^B+T_0({\bf x}_{k+1}
-{\bf x}_{k})\cdot\bgamma_{AB}{\tilde S}_k^B]
\end{equation}
Recalling the ``shortcut'' construction of the preceding
section, we see that we can modify these supercharges by
conjugating $Q_1+iQ_2$ with a nonunitary similarity
transformation of the form
\begin{equation}
V=\prod_kv(({\bf x}_{k+1}-{\bf x}_{k})^2, {\tilde S}_k)
\end{equation}
which will change the harmonic nearest neighbor potential to
an essentially arbitrary one, and at the same time introduce
terms into $Q_A$ cubic in $S, {\tilde S}$. This construction
then defines a supersymmetric chain dynamics with short-range
forces. The one particle quantum mechanics discussed
in this paper can be used in the approximate calculational scheme
developed in \cite{thornrpa}. The preservation of supersymmetry in
this approximation scheme will automatically enforce the
various subtractions necessary in passing to the continuum
limit.

The reader may well wonder why we are so concerned to have short
range bonding forces. After all, the continuum limit of the string-bit
polymers should wash out the details of the bonding force, so
why not be content with a harmonic one? This would probably
be a satisfactory position if we were only interested in
understanding free superstring and its {\it perturbative}
interactions. But our real aim is a bit higher, namely to
provide a physically sound basis for superstring theory
using string-bits as building blocks. In other words, we
want the dynamics of string-bits themselves to be physically
sensible. From a non-perturbative point of view a bit in
one piece of superstring can interact directly with one
in another piece of superstring: a model of polymer
bound states based solely on nearest neighbor interactions is
an approximate description, albeit one that can be singled
out by, for example, a $1/N_c$ expansion. A harmonic force
between nearest neighbor bits in a polymer would also, in
higher approximations, be present between non-neighbor
bits, including bits on different polymers. This would
strongly violate one of the most fundamental physical
properties of our world, cluster decomposition. While it
is barely conceivable that delicate cancellations could
be arranged to skirt this disaster, we think a much more satisfactory
and robust resolution of the difficulty
is to forbid the presence of such forces
in the fundamental dynamics from the beginning.
But then we are faced with the problems struggled with in this
paper.

What can be said about our inability to extend our construction
to $d=8$, the critical dimension for superstring? Perhaps
our range of search was too narrow. To broaden it we would have
to allow the momentum to enter the supercharges in a more
complicated way. Unfortunately, this greatly increases the
technical complications in enforcing the superalgebra. Also
one would have to guard against merely reproducing, after
much labor, the
trivial representation of the superalgebra described in
the introduction. Using the methods of this paper, however,
we can easily set up a dynamics with the degrees of freedom necessary
for critical superstring, but with only that part of
supersymmetry associated with the $d=2$ subspace realized.
This might be completely satisfactory. After all, our
physical world exists in 4 dimensional space-time, which
corresponds to this $d=2$ subspace. The practical virtues
of supersymmetry, namely an energy spectrum
bounded from below and the enforcement of necessary
cancellations, will be retained with such a partial
realization of supersymmetry. Finally, any supersymmetry
that remains has to be broken to account for the absence
of supersymmetry in our world. Perhaps it is not such a
tragedy if most of the supersymmetry of perturbative
superstring were simply not present in the underlying
dynamics, but is rather an artifact of perturbation theory.

Let's spell out this possibility in a little more detail. To
focus on the supersymmetry we wish to preserve, cast the
$SO(8)$ superalgebra in the language of $U(1)\times SU(4)$
where the $U(1)$ factor describes rotations in the
$d=2$ plane, and $SU(4)\simeq SO(6)$ describes rotations
in the remaining 6 directions (see, for example, Chapter 11
of \cite{greensw}). Then the 8
hermitian supercharges $Q^{\dot a}$ are replaced by
four non-hermitian charges $Q^A$, where $A$ labels
the components of a ${\bf 4}$ representation of
$SU(4)$. Then $Q_A\equiv Q^{A\dagger}$ transforms as a ${\bf\bar4}$.
Then the $SO(8)$ superalgebra takes on the appearance
of an $N=4$ extended $SO(2)$ superalgebra:
\begin{equation}
\{Q^A,Q^B\}=0\qquad \{Q^A, Q_B\}=\delta^A_B H.
\label{su4alg}
\end{equation}
With harmonic nearest neighbor forces this algebra is
fully satisfied without difficulty. Modifying these forces to short
range via the device of a nonunitary similarity transformation
will only preserve the first of Eq.\ (\ref{su4alg}). But
breaking the $SU(4)$ internal symmetry by singling
out one direction, say $Q\equiv Q^1$, allows the
specification of a dynamics
\begin{equation}
H\equiv\{Q,Q^\dagger\}
\end{equation}
consistent with an $N=1$ $SO(2)$ supersymmetry. This would
be the only vestige of supersymmetry present in the
underlying dynamics.

Finally we must note that our task of incorporating
$d=2$ supersymmetry into string-bit models is still far from
completion. We have dealt with the dynamics of a
``bare polymer'' chain with only nearest neighbor interactions.
This approximation arises in an $N_c\to\infty$ limit of
a second-quantized description, in which the
fields are $N_c\times N_c$ matrices, as
discussed in \cite{bergmantbits}. We have not
dealt with the problem of extending
the supersymmetry we have developed
for the first-quantized chain Hamiltonian to the second quantized
Hamiltonian. This is a non-trivial task we have yet to tackle.
\appendix
\setcounter{equation}{0}
\section{The Case $\lowercase{d}=4$}
In this appendix we analyze the case of four dimensions which
was not included in the main discussion because
the supercharges must be nonhermitian, entailing several
differences in detail. In four dimensions the Dirac matrices
can be chosen to be
\begin{equation}
\bgamma=\pmatrix{{\bf 0}&i\bsigma\cr -i\bsigma&{\bf 0}\cr}
\qquad \gamma^4=\pmatrix{0&I\cr I&0\cr},
\end{equation}
where we use bold-face to denote the first three components
of a four-vector and plain type-face to denote
all four components:  $x=({\bf x}, x^4)$.
In this representation the chirality matrix is diagonal. We
anticipate that the Weyl restriction is necessary for
harmonic forces, so we let our spinor variables be two component
from the beginning, and consider the following forms for
supercharges:
\begin{eqnarray}
Q_{\dot a}&=&(p^4-i\bsigma\cdot{\bf p})_{\dot ac}S^c+
k(x^4-i\bsigma\cdot{\bf x})_{\dot ac}{\tilde S}^c\nonumber\\
Q_{\dot a}^\dagger&=&(p^4+i\bsigma\cdot{\bf p})_{c\dot a}S^{c\dagger}+
k(x^4+i\bsigma\cdot{\bf x})_{c\dot a}{\tilde S}^{c\dagger}
\end{eqnarray}
The required superalgebra is then
\begin{equation}
\{Q_{\dot a}, Q_{\dot b}\}=0\qquad\{Q_{\dot a}, Q_{\dot b}^\dagger\}=
4\delta_{\dot a\dot b}H.
\label{dfoursu}
\end{equation}
With harmonic forces, the first of Eq.\ (\ref{dfoursu}) follows as a
consequence of the following identity satisfied by Pauli matrices:
\begin{equation}
\bsigma_{\dot ac}\cdot\bsigma_{\dot bd}-\delta_{\dot ac}\delta_{\dot bd}
=2\sigma^2_{\dot a\dot b}\sigma^2_{cd}
\label{id1}
\end{equation}
and the fact that $\sigma^2$ is antisymmetric. The second of
Eq.\ (\ref{dfoursu}) requires a variant of this identity, easily derived
by using $\bsigma^T=-\sigma^2\bsigma\sigma^2$:
\begin{equation}
\bsigma_{\dot ac}\cdot\bsigma_{d\dot b}+\delta_{\dot ac}\delta_{d\dot b}
=2\delta_{\dot a\dot b}\delta_{cd}.
\label{id2}
\end{equation}
The resulting supersymmetric oscillator Hamiltonian is
\begin{equation}
H={1\over2}(p\cdot p+k^2x\cdot x)-{ik\over2}(S^c{\tilde S}^{c\dagger}
-{\tilde S}^cS^{c\dagger}),
\end{equation}
where $v\cdot v={\bf v}^2+(v^4)^2$ for a four-vector $v$.

Now we want to explore the possibility of replacing
the harmonic force by a short-range one. In line with our
main discussion we impose the {\it ansatz} that the momentum
dependence of the supercharges is untouched
\begin{equation}
Q_{\dot a}=(p^4-i\bsigma\cdot{\bf p})_{\dot ac}S^c+
{\hat Q}_{\dot a}({x}, {\tilde S}, {\tilde S}^\dagger, S, S^\dagger).
\end{equation}
Just as in the main text, look first at the terms in the anticommutators
linear in $p$. The first equation of Eq.\ (\ref{dfoursu}) implies
\begin{eqnarray}
\{S^{\dot a},{\hat Q}_{\dot b}\}
+\{S^{\dot b},{\hat Q}_{\dot a}\}&=&0\nonumber\\
\sigma^k_{\dot ac}\{S^{c},{\hat Q}_{\dot b}\}
         +\sigma^k_{\dot bc}\{S^{c},{\hat Q}_{\dot a}\}&=&0.
\end{eqnarray}
Putting ${\dot b}={\dot a}$,
multiplying the second equation by $\sigma^k_{d\dot a}$,
summing over $k$, and using Eq.\ (\ref{id2}) and the first equation, leads to
$\{S^{c},{\hat Q}_{\dot a}\}=0$,
{\it i.e.} ${\hat Q}_{\dot a}$ is independent
of $S^\dagger$. Next the second equation of Eq.\ (\ref{dfoursu}) implies
\begin{eqnarray}
\{S^{\dot a},{\hat Q}_{\dot b}^\dagger\}
       +\{S^{\dot b\dagger},{\hat Q}_{\dot a}\}
  &=&\delta_{\dot a\dot b}V^4\nonumber\\
-i\sigma^k_{\dot ac}\{S^{c},{\hat Q}_{\dot b}^\dagger\}
         +i\sigma^k_{c\dot b}\{S^{c\dagger},{\hat Q}_{\dot a}\}&=&
         \delta_{\dot a\dot b}V^k.
\end{eqnarray}
Putting ${\dot b}={\dot a}$, multiplying the second equation
by $\sigma^k_{\dot ad}$,
summing over $k$, and using Eq.\ (\ref{id1}), Eq.\ (\ref{id2}) and the
first equation, allows
the determination of $\{S^{c\dagger},{\hat Q}_{\dot a}\}$ in
terms of $V$. Repeating
these manipulations, multiplying
instead by $\sigma^k_{d\dot a}$ determines
$\{S^{c},{\hat Q}_{\dot a}^\dagger\}$ in terms of $V$. The results are
\begin{eqnarray}
\{S^{c\dagger},{\hat Q}_{\dot a}\}&=&{1\over2}\delta_{\dot ac}V^4
-{i\over2}\bsigma_{\dot ac}\cdot{\bf V}                \nonumber\\
\{S^{c},{\hat Q}_{\dot a}^\dagger\}&=&{1\over2}\delta_{\dot ac}V^4
+{i\over2}\bsigma_{c\dot a}\cdot{\bf V}.
\end{eqnarray}
Since the left hand sides of these two equations are hermitian conjugates
of one another, it follows that $V^4$, ${\bf V}$ are hermitian. Also the
l.h.s side of the first is independent of $S^\dagger$ and that of the
second is independent of $S$, it follows that $V^4$, ${\bf V}$ are
independent of both $S$ and $S^\dagger$, and that ${\hat Q}_{\dot a}$ is
at most linear in $S$. In summary, the $p$ dependent terms of the
superalgebra imply that ${Q}_{\dot a}$ has the form
\begin{eqnarray}
Q_{\dot a}&=&(p^4+A^4(x, {\tilde S}, {\tilde S}^\dagger)
-i\bsigma\cdot({\bf p}
+{\bf A}(x, {\tilde S}, {\tilde S}^\dagger))_{\dot ac}S^c\nonumber\\
& &\qquad\mbox{}+{M}_{\dot a}(x, {\tilde S}, {\tilde S}^\dagger),
\end{eqnarray}
a result entirely analogous to Eq.\ (\ref{simpleq1}).

Next we look at the momentum independent terms bilinear in $S, S^\dagger$.
Isolating the bilinear terms in $\{Q_{\dot a}, Q_{\dot b}^\dagger\}$
produces
\begin{equation}
-\Big\{i(\delta_{\dot ac}\sigma^k_{d\dot b}
+\sigma^k_{\dot ac}\delta_{d\dot b})[D^4,D^k]+
\sigma^k_{\dot ac}\sigma^l_{d\dot b}[D^k,D^l]\Big\}S^{d\dagger}S^c,
\end{equation}
where we have abbreviated $D\equiv p+A$.
We require two Fierz style identities
\begin{eqnarray}
\sigma^k_{\dot ac}\sigma^l_{d\dot b}-\sigma^l_{\dot ac}\sigma^k_{d\dot b}
&=&\delta_{\dot a\dot b}i\epsilon^{lkm}\sigma^m_{dc}
+i\epsilon^{lmk}\sigma^m_{\dot a\dot b}\delta_{dc}\nonumber\\
\delta_{\dot ac}\sigma^k_{d\dot b}+\sigma^k_{\dot ac}\delta_{d\dot b}
&=&\delta_{\dot a\dot b}\sigma^k_{dc}
+\sigma^k_{\dot a\dot b}\delta_{dc}.
\end{eqnarray}
The terms proportional to $\delta_{\dot a\dot b}$ are compatible
with the superalgebra, leaving the ones which must vanish:
\begin{equation}
-i\sigma^m_{\dot a\dot b}\delta_{dc}([D^4,D^m]
+{1\over2}\epsilon^{lmk}[D^k,D^l])=0.
\end{equation}
Since the three Pauli matrices are independent, the following three
constraints must hold
\begin{equation}
[D^4,D^m]+{1\over2}\epsilon^{lmk}[D^k,D^l]=0,
\label{d4susy1}
\end{equation}
which is just the statement that the antisymmetric ``field strength''
$F^{\mu\nu}$ is self dual. The analysis of the
bilinear tems in $\{Q_{\dot a}, Q_{\dot b}\}=0$
turns out to add no new constraints. The easiest way to see this
is to use in place of $Q_{\dot b}$ the related operators
\begin{eqnarray}
(\sigma^2Q)_{\dot b}
&=&(p^4+A^4(x, {\tilde S}, {\tilde S}^\dagger)\nonumber\\
& &\qquad\mbox{}
+i\bsigma\cdot({\bf p}
+{\bf A}(x, {\tilde S}, {\tilde S}^\dagger))_{d\dot b}(\sigma^2S)^d
\nonumber\\
& &\mbox{}\qquad\qquad\qquad\qquad
+(\sigma^2M)_{\dot b}(x, {\tilde S}, {\tilde S}^\dagger),
\end{eqnarray}
so the bilinear terms in $\{Q_{\dot a}, (\sigma^2Q)_{\dot b}\}$
are identical to those in $\{Q_{\dot a}, Q_{\dot b}^\dagger\}$ with
$S^{d\dagger}S^c$ replaced by $(\sigma^2S)^{d}S^c$. The one difference
is that the terms proportional to $\delta_{\dot a\dot b}$ must now also
vanish. But they do because those terms involve the factor
$\sigma^m_{dc}(\sigma^2S)^{d}S^c=-(\sigma^2\sigma^m)_{dc}S^dS^c$
which vanishes identically because $\sigma^2\sigma^m$ are
symmetric matrices. Thus we find that the case $d=4$ is sort
of intermediate between $d=2$ and $d=8$: some but not all
linear combinations of the components of $F$ must vanish:
$F$ must be self-dual.

Assuming self-duality the anticommutators of supercharges reduce to
\begin{eqnarray}
\{Q_{\dot a}, (\sigma^2Q)_{\dot b}\}&=&
[M_{\dot a},(D^4+i\bsigma\cdot{\bf D})_{d\dot b}](\sigma^2S)^{d}
+[(\sigma^2M)_{\dot b},(D^4-i\bsigma\cdot{\bf D})_{\dot ac}]S^{c}
+\{M_{\dot a}, (\sigma^2M)_{\dot b}\}=0\nonumber\\
\{Q_{\dot a}, Q_{\dot b}^\dagger\}&=&((D^4)^2
+{\bf D}^2)\delta_{\dot a\dot b}
+[M_{\dot a},(D^4+i\bsigma\cdot{\bf D})_{d\dot b}]S^{d\dagger}\nonumber\\
& &\qquad\qquad\qquad\qquad\qquad
\mbox{}+[M_{\dot b}^\dagger,(D^4-i\bsigma\cdot{\bf D})_{\dot ac}]S^{c}
+\{M_{\dot a}, M_{\dot b}^\dagger\}
=4\delta_{\dot a\dot b}H.
\end{eqnarray}
\noindent The linear terms in $S,S^\dagger$ must separately be proportional
to $\delta_{\dot a\dot b}$ in the second equation and must vanish
in the first equation. From the second equation we conclude that
\begin{equation}
[M_{\dot a},(D^4+i\bsigma\cdot{\bf D})_{d\dot b}]=
\Psi_d\delta_{\dot a\dot b},
\label{d4susy2}
\end{equation}
which implies also its hermitian conjugate. Once this holds the
corresponding linear terms in the first equation automatically
vanish, as can be easily shown. Finally the terms independent of
$S$ imply
\begin{equation}
\{M_{\dot a}, M_{\dot b}^\dagger\}=4V\delta_{\dot a\dot b}
\qquad\{M_{\dot a}, M_{\dot b}\}=0.
\label{d4susy3}
\end{equation}

In summary, supersymmetric quantum mechanics in $d=4$ will be realized
if solutions to Eq.\ (\ref{d4susy1}), Eq.\ (\ref{d4susy2}),
and Eq.\ (\ref{d4susy3})
can be found. In these equations $M$ and $A$ are allowed to depend only
on the coordinates ${\bf x}$ and on the fermionic variables
${\tilde S},{\tilde S}^\dagger$. We defer the search for such
solutions and a study of their properties for a later time.



\end{document}